\documentclass[journal,transmag]{IEEEtran}

\usepackage{graphicx}
\usepackage{cite}
\usepackage{float}
\usepackage{subcaption}
\usepackage{listings}
\usepackage{subcaption}
\usepackage{balance}
\usepackage{hyperref}
\hypersetup{
  pdfinfo={
  pdfproducer={},
  Title={},
  Subject={},
  Author={},
  }
}

\usepackage{color}

\definecolor{dkgreen}{rgb}{0,0.6,0}
\definecolor{gray}{rgb}{0.5,0.5,0.5}
\definecolor{mauve}{rgb}{0.58,0,0.82}

\ifCLASSINFOpdf
\else
\fi
\begin{document}
%
\title{Pay as You Go: A Generic Crypto Tolling Architecture}


\author{\IEEEauthorblockN{Paulo C. Bartolomeu, Emanuel Vieira, and Joaquim Ferreira}
\IEEEauthorblockA{Institute of Telecommunications, University of Aveiro}}

%



\IEEEtitleabstractindextext{%
\begin{abstract}
The imminent pervasive adoption of vehicular communication, based on dedicated short-range technology (ETSI ITS G5 or IEEE WAVE), 5G, or both, will foster a richer service ecosystem for vehicular applications. The appearance of new cryptography based solutions envisaging digital identity and currency exchange are set to stem new approaches for existing and future challenges. This paper presents a novel tolling architecture that harnesses the availability of 5G C-V2X connectivity for open road tolling using smartphones, IOTA as the digital currency and Hyperledger Indy for identity validation. An experimental feasibility analysis is used to validate the proposed architecture for secure, private and convenient electronic toll payment.

\end{abstract}

}

\maketitle

\IEEEdisplaynontitleabstractindextext

%
\IEEEpeerreviewmaketitle

\section{Introduction}

Recent advances in wireless communications and crypto technologies are fostering the emergence of innovative solutions for cooperative connected and automated mobility (CCAM). In CCAM, the road infrastructure plays a central role in providing collaborative awareness data to connected and automated vehicles and other road users. 
Dependable vehicular communications are a cornerstone of CCAM. For more than ten years, research and development in this area have produced mature technology ready to me massively deployed, notably ETSI ITS G5 and IEEE WAVE, which share the same physical and MAC layers, initially adapted from IEEE 801.11. Recently, however, the emergence of the fifth generation of cellular mobile communications (5G) and other related cellular and device-to-device (D2D) technologies, notably cellular Vehicle to everything (C-V2X), have raised some doubts on which vehicular communication technology will prevail. 

On the other side, recent advances in cryptography have also led to the appearance of new digital currencies and novel identity paradigms, all of which will be crucial for enabling the future of digital transactions among people, organizations and things. 

An arena where such novelty combination will have a tremendous impact is on vehicular applications. The impending evolution of autonomous electric vehicles is expected to stem multiple services that will enhance the passengers' user experience and provide better returns for providers as never before was possible. One service that can largely benefit from these innovations is the automatic free-flow tolling of vehicles in licensed roads, for which existing solutions are not able to meet stringent privacy requirements and provide real-time operation.

Road tolling, as a method of financing the transportation system, has long been in place all around the world. Since neither the drivers or the road operators want vehicles to stop or slow down to pay to use a toll road, several technologies, collectively called Electronic Toll Collection (ETC), have been developed in the last 25 years, ranging from RFID sticker toll tags, to Dedicated Short Range Communications (DSRC) and to tolling systems based on an autonomous On Board Unit (OBU) using Global Navigation Satellite System/Cellular Network (GNSS/CN). Currently, Open Free-Flow Road Tolling (ORT), with all-electronic toll collection, is the preferred practice, as it is more environmentally friendly, and safer than manual toll collection \cite{Persad07}. Electronic tolling is cheaper than a staffed booth, reducing the average cost per transaction. With electronic tolling one can also vary the amount of the toll, implement road congestion pricing, including for high-occupancy lanes, toll lanes that bypass congestion, and city-wide congestion charges. 

Open road tolling systems need to classify vehicles, so different vehicles types can pay be charged different rates. Automated vehicle identification (AVI) systems use a variety of sensors for vehicle classification, including inductive loops to count the number of axles and light-curtain laser profilers to detect the shape of the vehicle, thus helping distinguishing trucks and trailers and to measure their height.

Violation enforcement systems need to be put in place together with open road tolling to minimize unpaid tolls. To this end, several technologies can be adopted, possibly including physical barriers or just automatic number plate recognition systems synchronized with the on board transponder detection. 

Transaction processing, a core function of the electronic toll collection, is responsible for charging customer accounts. Customer accounts can be postpaid, where toll transactions are periodically billed to the customer, or prepaid, where the customer funds a balance in the account which is then depleted as toll transactions occur. Depending on the country and on the adopted payment infrastructure, it can take several days to charge the customer account, for the case of the postpaid micro-payments.   

Traditional open free-flow road tolling systems usually have a national or regional scope, making it cumbersome for unregistered drivers, e.g., tourists, to pay the toll, as there might not be toll booths with alternative payment methods. Additionally, these systems require the installation of a device in each vehicle, complex toll gantries and a centralized payment processing system.

This paper proposes a generic tolling architecture to overcome some of the issues of current open free-flow road tolling systems. The new tolling architecture, is leveraged by the IOTA cryptocurrency, for decentralized fee-less payments, the Hyperledger Indy, for decentralized self-sovereign identity trust, and C-V2X for connectivity with the road side infrastructure. Furthermore, the proposed system does not require specialized devices installed in vehicles. 
The rest of the paper is organized as follows: section \ref{key-techs} provides a brief overview of the key technologies required to build the new tolling system, while section \ref{crypto-toll-arch} presents the novel tolling architecture, which feasibility is experimentally evaluated in section \ref{feasibility}. Section \ref{rel-work} presents an overview of related work in the area of secure 5G communications and, finally, section \ref{conclusions} summarises the contributions of the paper.

\section{Key Emerging Technologies}
\label{key-techs}

The future of open road tolling systems will be shaped by emerging technologies such as 5G communications, new crypto currencies and distributed ledger identity networks. In this section a brief overview of selected technologies and their potential impact on electronic tolling systems is provided.
Although it is possible to use ETSI ITS G5 or IEEE WAVE communications to convey the tolling transactions between vehicles and the road side infrastructure, the crypto tolling architecture proposed in this paper considers the use of C-V2X communications, mainly to allow the use of smartphones instead of specialized devices installed in the vehicles. Notice, however, that it is still not clear whether the smartphones' chipsets will support C-V2X. If not, the proposed architecture can seamlessly be adapted to work with the C-V2X on board unit (OBU).

\subsection{C-V2X communications}
\label{5g-d2d}

Connectivity is the touchstone of the Internet of Things (IoT) era. The 5G promise of pervasive high bandwidth and low-cost wireless communications suggests a bright future for IoT, as an heterogeneous  range of applications will emerge under this umbrella. Scenarios where services are enabled by the occurrence of direct (and automated) interactions among ``things" represent one of the most interesting opportunities for the 5G Device-to-Device technology. Indeed, such scenarios allow alleviating the traffic load from base stations by taking advantage of the proximity among devices, which not only contributes to improve parameters such as throughput, latency and power usage, but also increases the overall capacity of the network \cite{vlachos2017} and increases its reliability.

The 3GPP Release 14 introduced the C-V2X standard \cite{cv2x}, also known as LTE-V or LTE-V2X, that uses the LTE PC5 interface for vehicle-to-vehicle (V2V) communications.
This standard, designed to support both cooperative traffic safety and efficiency applications, includes two modes of operation, namely C-V2X mode 3 and C-V2X mode 4. 
In C-V2X mode 3, scheduling and interference management of V2V traffic are assisted by eNBs via control signaling, while in C-V2X mode 4 scheduling and interference management of V2V traffic are based on distributed algorithms implemented between UEs. 
C-V2X mode 3, can thus be used for long-range network communications (V2N), relying on the conventional mobile network to enable a vehicle to receive cooperative perception data originated beyond the vehicle´s line of sight. 
On the other side, C-V2X mode 4 is used for short-range direct communications between vehicles (V2V), between vehicles and infrastructure (V2I), and vehicles and other road users (V2P), such as cyclists and pedestrians. C-V2X mode 4 works independently of the cellular networks, in dedicated ITS 5.9GHz spectrum.


Considering an open road tolling system as the target application, this paper proposes the use of C-V2X mode 4, as it can support safety applications in the absence of coverage from the cellular infrastructure, which might not be present in remote areas. 

Concerning the security aspects of C-V2X, the 3GPP R14 specifications \cite{cv2x} state that no security and privacy is applied for the PC5 broadcast communication by setting the fields related to group security to 0. In this way, messages exchanged among UEs have no standard security mechanisms in place. Application layer security is outside the scope of 3GPP, but it is suggested that credentials should be periodically refreshed to avoid UE tracking. Security and privacy mechanisms for C-V2X mode 4 need to be defined by regulators and operators.
As for authentication, traditional LTE authentication mechanisms can be employed and enforced in C-V2X mode 3. But, since an operator may not be present in mode 4, proper authentication has to rely on other schemes.

\subsection{IOTA}
\label{iota}

IOTA is a cryptocurrency that was created with a focus on IoT to solve the problems of scalability, control centralization, transaction fees and post-quantum security that characterizes other cryptocurrencies employing the blockchain technology \cite{e1}. The \emph{Tangle} is its key contribution and builds on the concept of Directed Acyclic Graphs (DAGs) as a substitute for blockchains. 

In the \emph{Tangle} every vertex is a transaction. To add a transaction to the \emph{Tangle} two other transactions must be approved. This approval can be represented by an outgoing edge, meaning that the outdegree of every vertex is equal to 2. The indegree is not limited, a transaction can be approved by any number of transactions. Transactions which are currently not approved by any others are called tips. Similarly to blockchains, for a transaction to be appended to the \emph{Tangle}, some ``proof-of-work" (PoW) must be executed, mainly to avoid spam attacks.

\subsection{Hyperledger Indy}
\label{hyperledger-indy}
The Hyperledger Indy is a distributed ledger built for decentralized identities. It uses a blockchain as a ledger and employs a permissioned protocol where only trusted elements can add transactions to the ledger. The trust factor removes the need for users to do any PoW, shortening the time required to add a transaction. Whenever a transaction needs to be updated, a new updated one is added to the ledger. The older transaction, whilst still existing in the ledger, is simply ignored in future queries. 

Hyperledger Indy is mostly known for its main implementer, the Sovrin Foundation. Sovrin is a global trust network that provides the legal and trust foundation for self-sovereign identities (SSI) \cite{sovrinwhitepaper}. Four roles with specific scopes and permissions are defined: Trustees, Stewards, Trust Anchors and Identity Owners. In Hyperledger Indy any user is able to issue its own credentials. An issued credential is stored in the user's wallet and can be provided to prove certain attributes to a proof requester, also called \textit{Verifier}. Every created proof is signed by the credential \textit{Issuer}. If the \textit{Verifier} trusts the \textit{Issuer} then he can also trust the created proof.

\section{Crypto Tolling}
\label{crypto-toll-arch}

The proposed open road tolling architecture arises from the opportunity to adopt a decentralized approach for identification and payment that ensures adequate anonymity and, at the same time, provides accountability and traceability. The modern smartphone is established at the center of the ``Pay as You Go" architecture. This device has the required computational power to run blockchain wallets, connectivity to communicate via 5G D2D and a GPS to monitor its physical location and assist in the validation of toll payments.

There are several motivations to adopt a distributed SSI mechanism enabling devices to prove their identities directly to each other without a central authority involved.  Devices are closer to each other during information exchange (identity validation) making the process typically faster. Because the data exposure is set to the minimum required to fulfill a given function, personal data has an higher level of protection. The information controlability is also higher because the user/device is in control of the information that it shares. Finally, because the identity information is distributed among wallets located in devices that can be mobile (e.g., smartphones) there is also an higher level of portability. 

The adoption of a decentralized payment approach was driven by several factors. IOTA has ultra low fees (or no fees at all) applicable to transactions, which is a strong enabler for micro payment systems, such as open road tolling. Because IOTA is a globally available cryptocurrency it can be traded worldwide without control of central authorities/governments. Finally, when compared to other cryptocurrencies such as Bitcoin, IOTA transactions are confirmed much faster, thus allowing payments to be available to the receiving party much earlier.

Although the use case presented in this paper specifically addresses the open road toll payment use case, the proposed architecture and its operation can be easily generalized to other automatic service or product payments that can work under similar requirements.

\subsection{Architecture}
\label{arch}

The ``Pay as You Go" architecture comprehends several stakeholders: \textit{Road Operator}, \textit{Toll Gantry}, \textit{Vehicle} and \textit{User}. The \textit{Operator} is generally the company who has the concession of a given highway and charges the \textit{Vehicle}'s owner for its use. The \textit{Operator} does so by employing automatic \textit{Toll Gantries} that are used to register information about \textit{Vehicles} that pass by. This information is then used to either automatically collect payment from the \textit{Vehicles}' owners bank account or to provide mechanisms for posterior enforcing of voluntary or forceful charging. In both cases, images of the vehicle are recorded for traceability. \textit{Vehicles} have distinct physical attributes (licence plate, toll class, etc.) which allow their identification in an unambiguous way. Finally, \textit{Users} that possess a valid driving licence are allowed to drive their \textit{Vehicles} on roads covered by \textit{Toll Gantries}, whereas the driver' smartphone is the center element for paying the toll service.

\begin{figure}[t]
	\centering
	\includegraphics[width=0.5\textwidth]{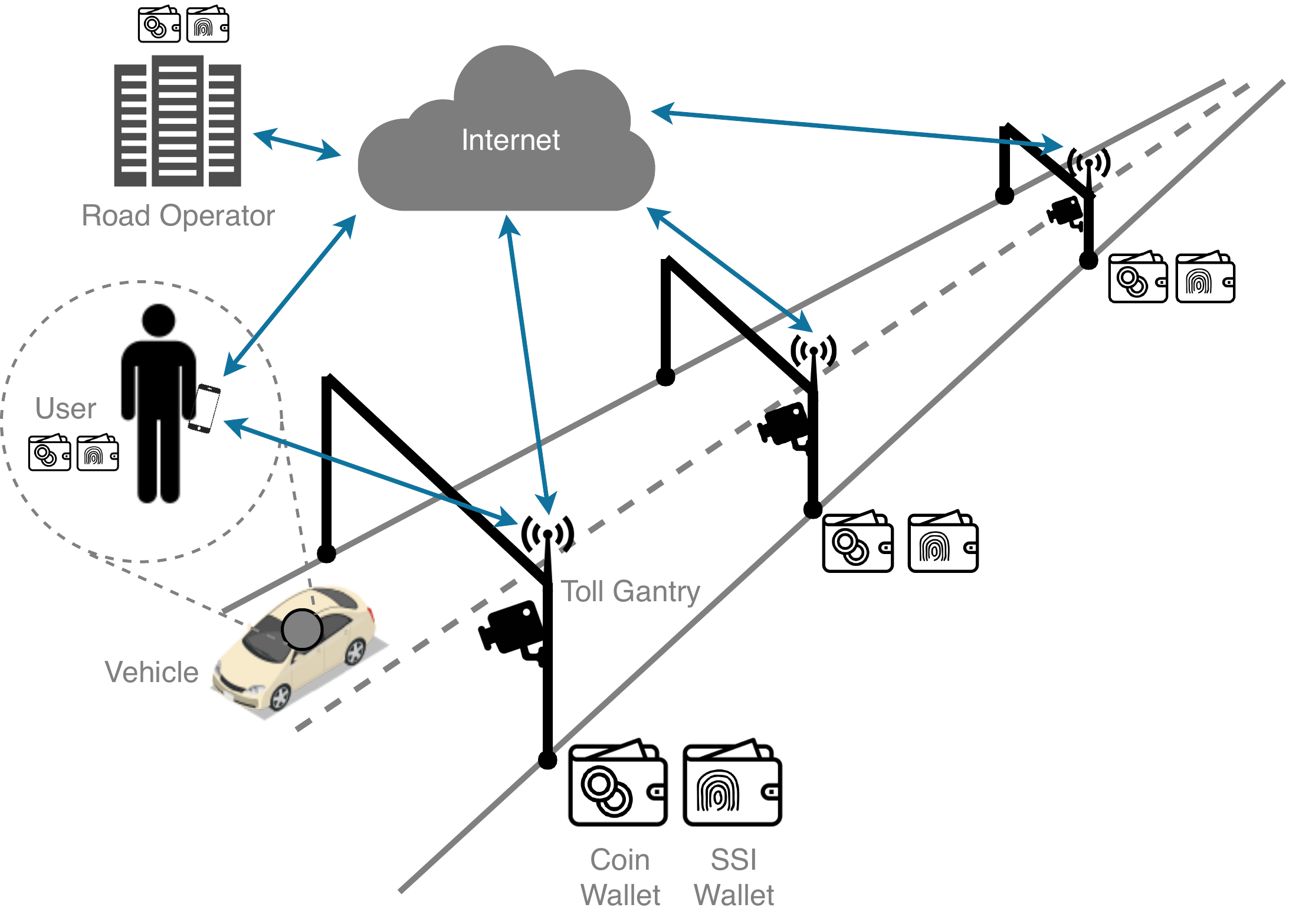}
        \caption{``Pay as You Go" architecture.}
        \label{fig:paygo-arch}
\end{figure}

As depicted in Fig. \ref{fig:paygo-arch}, all stakeholders are interconnected. it is assumed that the \textit{User} smartphone is equipped with 5G and communicates with the \textit{Toll Gantries} using C-V2X mode 4. The \textit{Road Operator} can communicate with the \textit{Toll Gantries} via a wired/fiber backhaul connection to the Internet or, also, via 5G. 

Each of the participating entities possesses a set of unique identifications saved in a  wallet that is used to authenticate them and prove their veracity. Besides the identity wallet, entities will have coin wallets which are used to make and receive IOTA payments.

\subsection{Operation}
\label{operation}

The proposed crypto architecture relies on three key elements: 5G C-V2X mode 4 connections between the \textit{User}'s smartphone and the tolling infrastructure, Hyperledger Indy for identity management/validation and IOTA currency to conduct the tolling service micro payments. 

In the context of this paper, it is assumed that 5G C-V2X mode 4 communications will be seamlessly established between the \textit{User}'s smartphone and the \textit{Toll Gantries}. In this sense, the trigger mechanism for initiating a new toll payment can be implemented in multiple ways. For example, when the connection between devices is physically established, it can redirect the \textit{User}'s smartphone to a Web page where it automatically gets an invite to establish a connection, in a similar way to what is typically done by companies offering free Wi-Fi. Another option is having an App in the mobile phone that listens to specific ``out-of-band" announcements broadcasted by the \textit{Toll Gantries} and, when such a request is detected, the smartphone is triggered to initiate an ``in-band" connection to the \textit{Toll Gantry}.

The smartphone App will perform an early check to evaluate if there is a previous ongoing digital relationship with the \textit{Toll Gantry} in range before attempting to create a new pairwise DID for the relation. This evaluation can be realized by storing in the App the relationship between known \textit{Toll Gantry} DIDs and their geographical location (as perceived in previous interactions). In this case, when a new interaction is about to begin, the smartphone App reads its geographical localization from the GPS and filters the list of possible DIDs, selecting the one that is closer (minimal localization error). Then, the smartphone App (re)establishes the pairwise DID connection by echoing a request and receiving a response from the \textit{Toll Gantry}.

The creation of verinyms (DIDs) associated with the identities of the \textit{User}, \textit{Toll Gantry}, \textit{Operator}, etc. is beyond the scope of this paper. The considered scenario assumes that the edge wallets of the \textit{User} and \textit{Toll Gantry} have already been populated with the necessary credentials generated by third party trusted organizations or authorities that are recognized to provide them. In this context, it is also assumed that both parties have already created their \textit{Master Secrets} allowing to guarantee that a given credential uniquely applies to them.

\subsubsection{Pairwise DID creation}

\begin{figure}[t]
	\centering
	\includegraphics[width=0.50\textwidth]{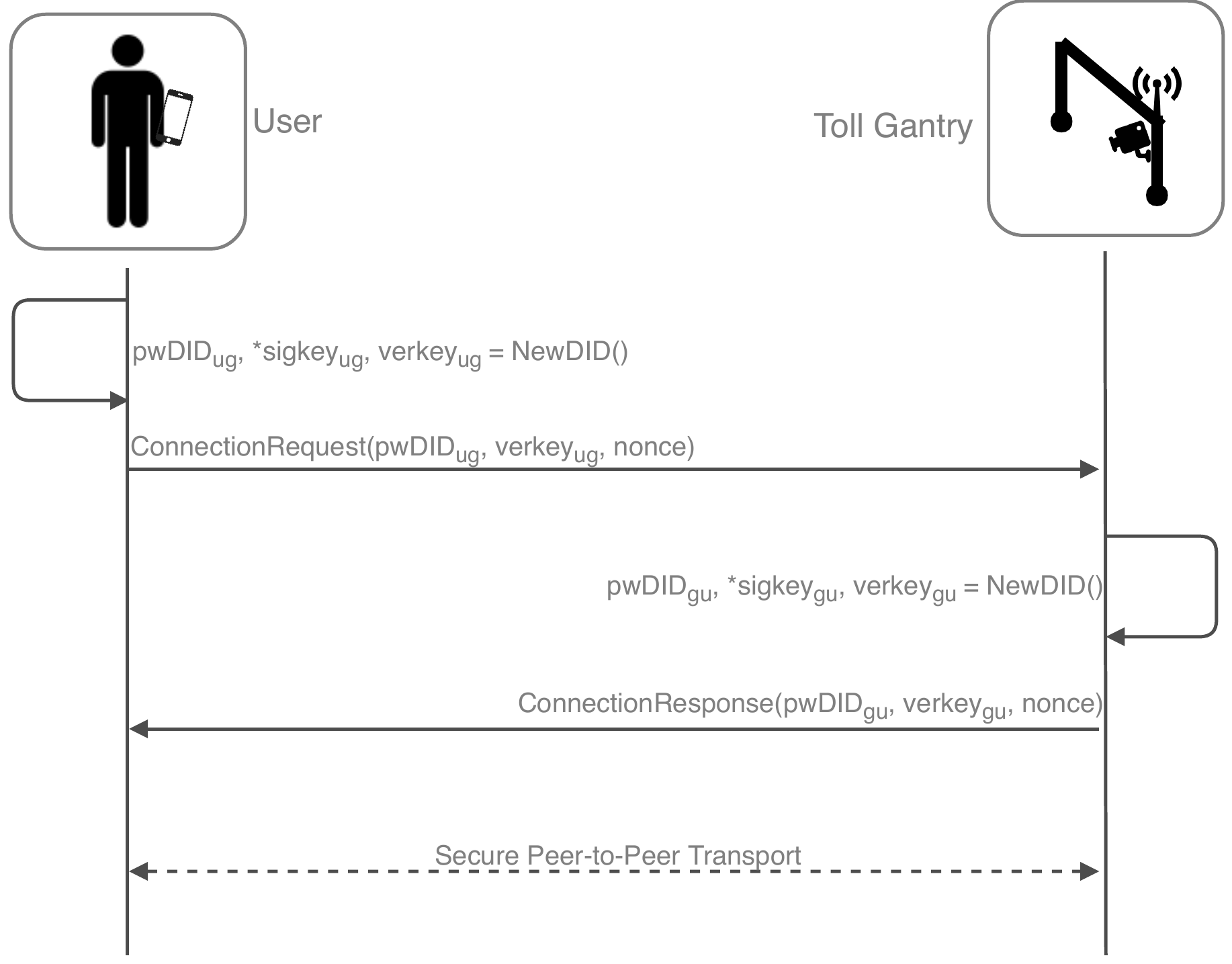}
        \caption{Establishing a pairwise DID transport}
        \label{fig:indy-pairwise-connection}
\end{figure}

If the connection is successfully re-established, then a new exchange and validation of credentials can begin. Otherwise, a new pairwise DID relation is required to be established following the diagram of Fig. \ref{fig:indy-pairwise-connection}. As documented, the \textit{User} agent running in the smartphone begins by creating a DID and associated keys and storing them in its local wallet. This new DID will be used only in the context of the interactions with this \textit{Toll Gantry}. The impossibility of reading the signing (private) key ($sigkey_{ug}$) stored in the wallet is distinguishable by the use of an asterisk in its notation.

Subsequently, the \textit{User} agent sends a connection request to the \textit{Toll Gantry} agent encompassing the created DID, verification key and nonce. The verification key allows the \textit{Toll Gantry} to check the authenticity of the message, i.e., that it was sent by the \textit{User}. The nonce is used by the initiating party (in this case the \textit{User} agent) to correlate the response with the request. 

Upon receiving the connection request, the \textit{Toll Gantry} agent also creates a new DID to be used in the context of this unique digital relationship. Afterwards, it sends an encrypted connection response using the \textit{User}'s verification key (\textit{verkey}) that encompasses the newly created DID, verification key and the nonce originally received from the \textit{User} agent. Because messages in both directions can now be encrypted using the verkeys of the target party, a unique secure peer-to-peer transport is established between \textit{User} and \textit{Toll Gantry}.

\subsubsection{Credential validation}
\begin{figure}[t]
	\centering
	\includegraphics[width=0.50\textwidth]{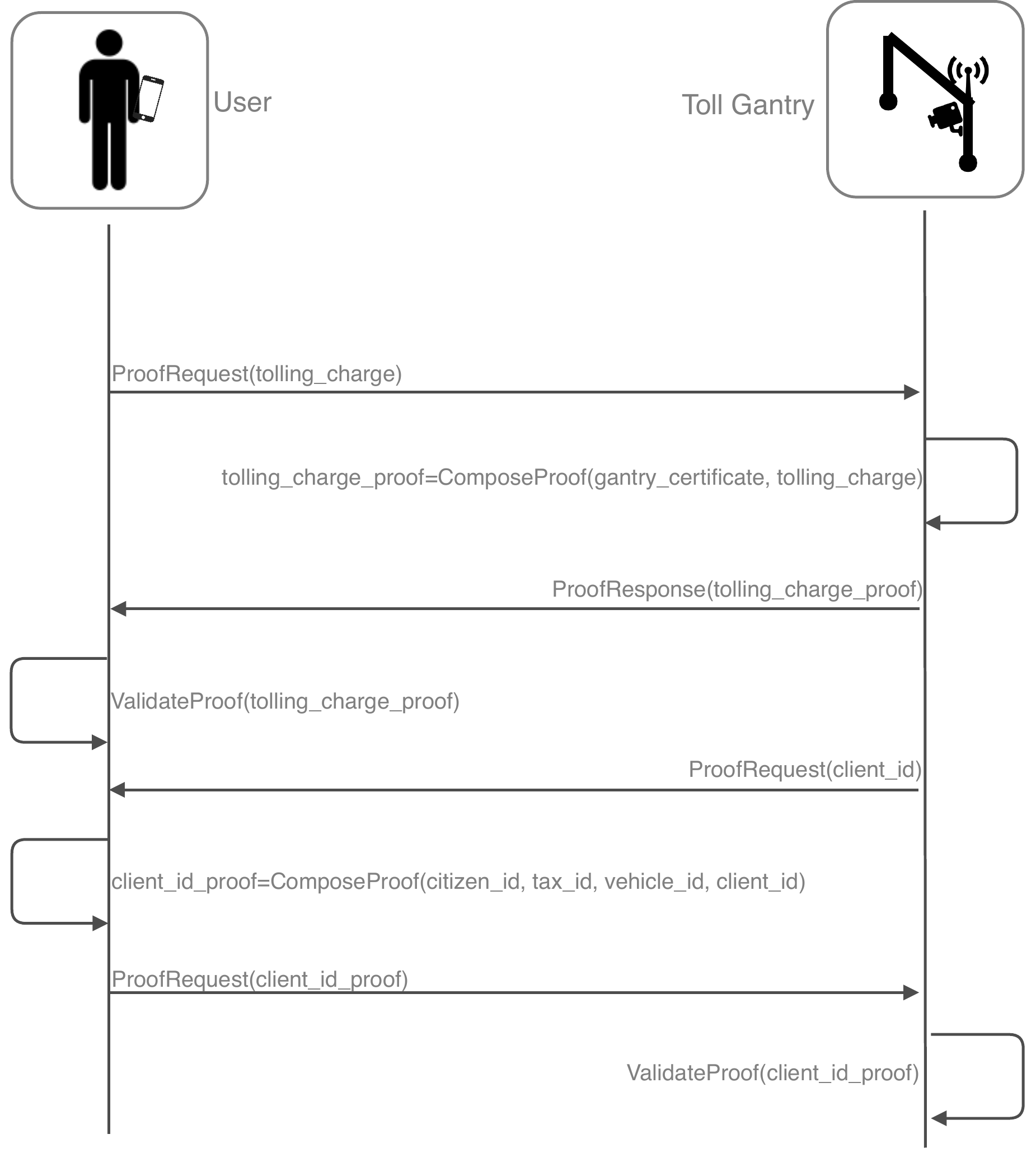}
        \caption{\textit{User} and \textit{Toll Gantry} identity validation}
        \label{fig:indy-proof-cycle}
\end{figure}

The credentials validation phase is stemmed by the \textit{User} edge agent that sends a \textit{Tolling Charge} proof request to the \textit{Toll Gantry} in order to collect verifiable proof that its attributes meet specific criteria. In this case, the \textit{Tolling Charge} requires a name, localization and unique identifier code. All of these credential parameters must be formally asserted by a tolling operator. As depicted in Fig. \ref{fig:indy-proof-cycle}, the \textit{Toll Gantry} agent then composes a proof based on the received proof request and on the \textit{Toll Gantry} certificate that is stored on its wallet.

On the opposite direction, the \textit{Toll Gantry} agent sends a \textit{Client Identification} proof request requiring the \textit{User}'s name, VAT number and vehicle plate number. The exposed information about the \textit{User} is required for tax purposes when the payment is issued (e.g., to issue an invoice), while  the information about the vehicle is used in its validation.

The direction of the vehicle passage can easily be detected by the cameras of the enforcement system that capture its passage.

\subsubsection{Payment}
After the \textit{User} and \textit{Toll Gantry} identities are validated, the \textit{Toll Gantry} sends the payment request to the \textit{User}'s smartphone. This request contains the due amount, the IOTA address to transfers IOTAs to, and a nonce for the \textit{User} to use as a reference. This bill will be sent encrypted using the pairwise DID created by the \textit{User}. 

The tolling App can then use this information to proceed with the IOTA transaction that transfers the due amount to the designated address. If the payment with the specific nonce and value is not received within a pre-defined period of time in the designated address, the operator may trigger a legal process using the information collected in the enforcement system (license plate number, vehicle characteristics, photos, etc.) and, if available, information obtained during identity validation.

\section{Feasibility Evaluation}
\label{feasibility}

This section evaluates the feasibility of implementing the proposed payment architecture with the constrains imposed by its operation. For this analysis a C-V2X mode 4 line of sight communication range of 600  meters \cite{5gaa} and a maximum speed of 130 Km/h were considered. The speed corresponds to the legal limit that covers most of the countries around the world. The identity ``handshake" between the \textit{User}'s smartphone and the \textit{Toll Gantry} plus the IOTA value transaction registration into the \textit{Tangle} needs to be completed during the period in which the smartphone is in range of the \textit{Toll Gantry}, i.e., 33.2 seconds considering a coverage of 1.2 Km (twice the communication range). 

In this evaluation, given the unavailability of 5G terminal equipment, it was decided to use 10/100 Ethernet as the communication technology to simulate the C-V2X mode 4 communication between \textit{User} device and \textit{Toll Gantry}. 

In the following subsections, the setup and timing results obtained for both \textit{Credential Validation} and \textit{IOTA Payment} are presented and discussed.

\subsection{Credential Validation}

\begin{figure*}
\centering
\begin{subfigure}{0.6\textwidth}
  \centering
  \includegraphics[width=0.65\linewidth]{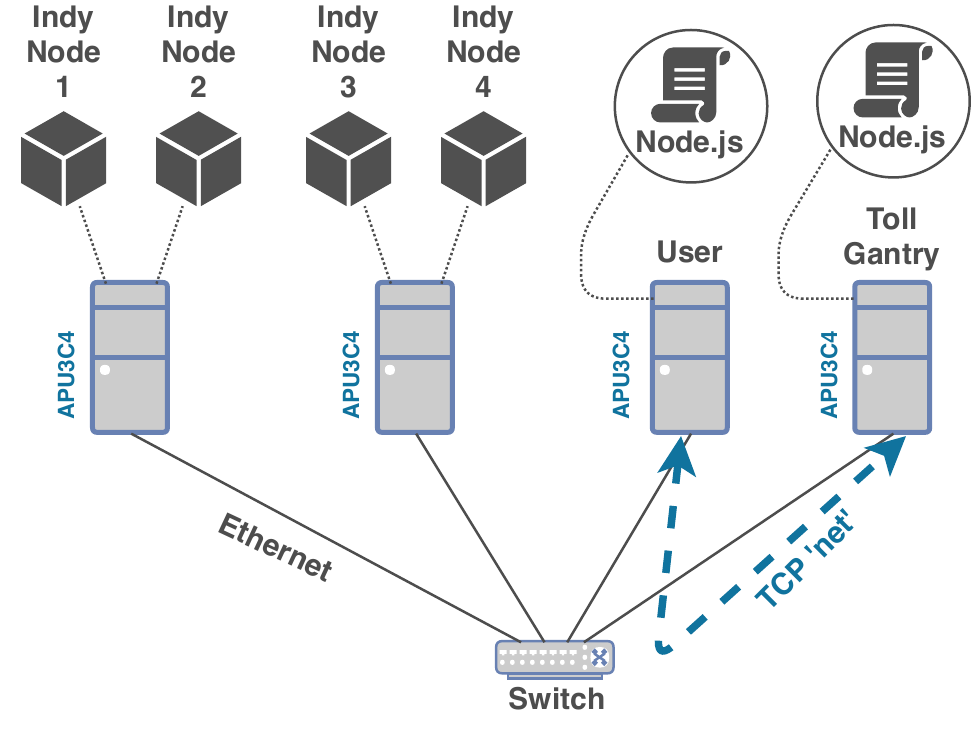}
  \caption{Hyperlegder Indy evaluation setup.}
  \label{fig:sub:indy}
\end{subfigure}%
\begin{subfigure}{0.4\textwidth}
  \centering
  \includegraphics[width=1\linewidth]{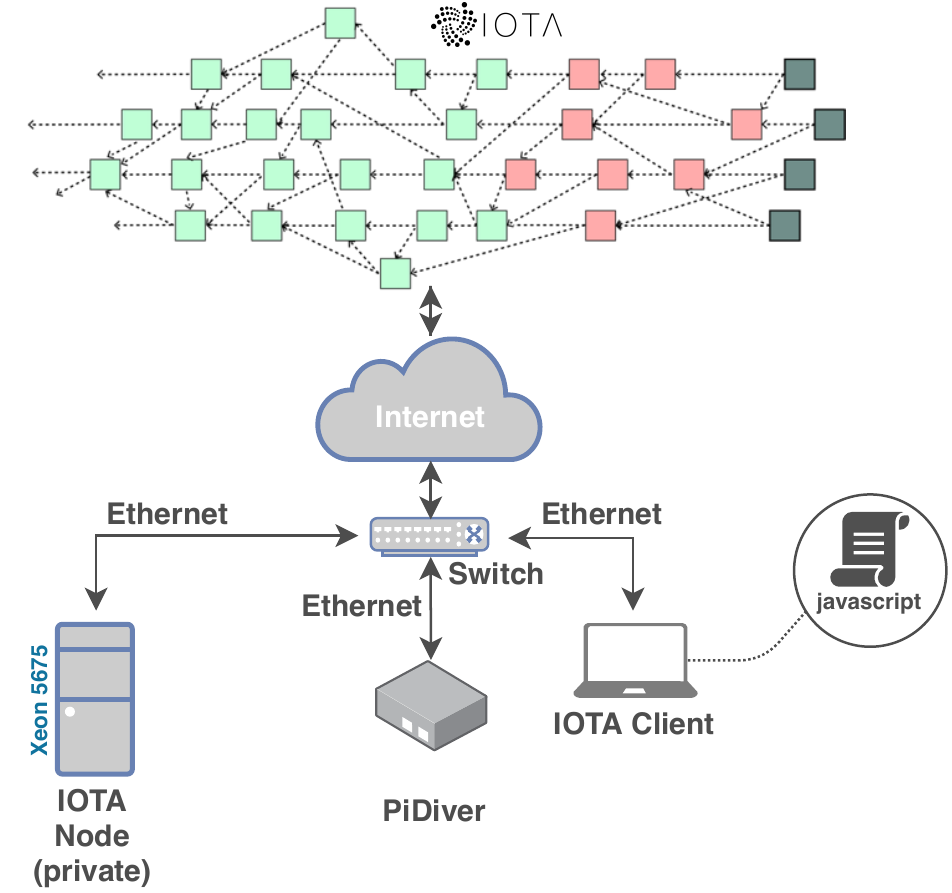}
  \caption{IOTA evaluation setup.}
  \label{fig:sub:iota}
\end{subfigure}
\caption{Experimental setups used to evaluate the latency of performing credential validations with the Hyperledger Indy framework and conducting digital money transactions through the IOTA network.}
\label{fig:eval-setup}
\end{figure*}

The experimental setup shown in Fig. \ref{fig:sub:indy} was realized to evaluate the credential validations between \textit{User} and \textit{Toll Gantry}. For this purpose a set of four APU3C4 embedded PCs running Arch Linux was used. Each PC encompasses a quadcore AMD Jaguar processor clocked at 1 GHz and 4 GB of RAM memory. Two PCs were setup to run a total of four Hyperledger Indy nodes while the other two were used as Identity Owners, one representing the \textit{User} and the other one acting as the \textit{Toll Gantry}. 

The Indy node pool was setup using Docker together with version 1.6.78 of the Hyperledger Indy. On the Identity Owner side, the Node.js wrapper for the Indy SDK version 1.6.7 was employed. The \textit{Toll Gantry} and \textit{User} interaction was simulated with  two different Node.js scripts built upon the TCP 'net' module. A basic communication performance test with the mentioned 'net' module allowed to conclude that the average round trip delay between \textit{Toll Gantry} and \textit{User} PC was 70 ms.

In order to assess the time required to validate the identities of the \textit{Toll Gantry} and of the \textit{User} in an Hyperledger Indy setup, following the interaction sequences depicted in Figs. \ref{fig:indy-pairwise-connection} and \ref{fig:indy-proof-cycle}, a set of 1000 latency measurements were conducted for the process of establishing a pairwise DID transport plus \textit{User} and \textit{Toll Gantry} identity validation. 

\begin{figure*}[t]
	\centering
	\includegraphics[width=0.70\textwidth]{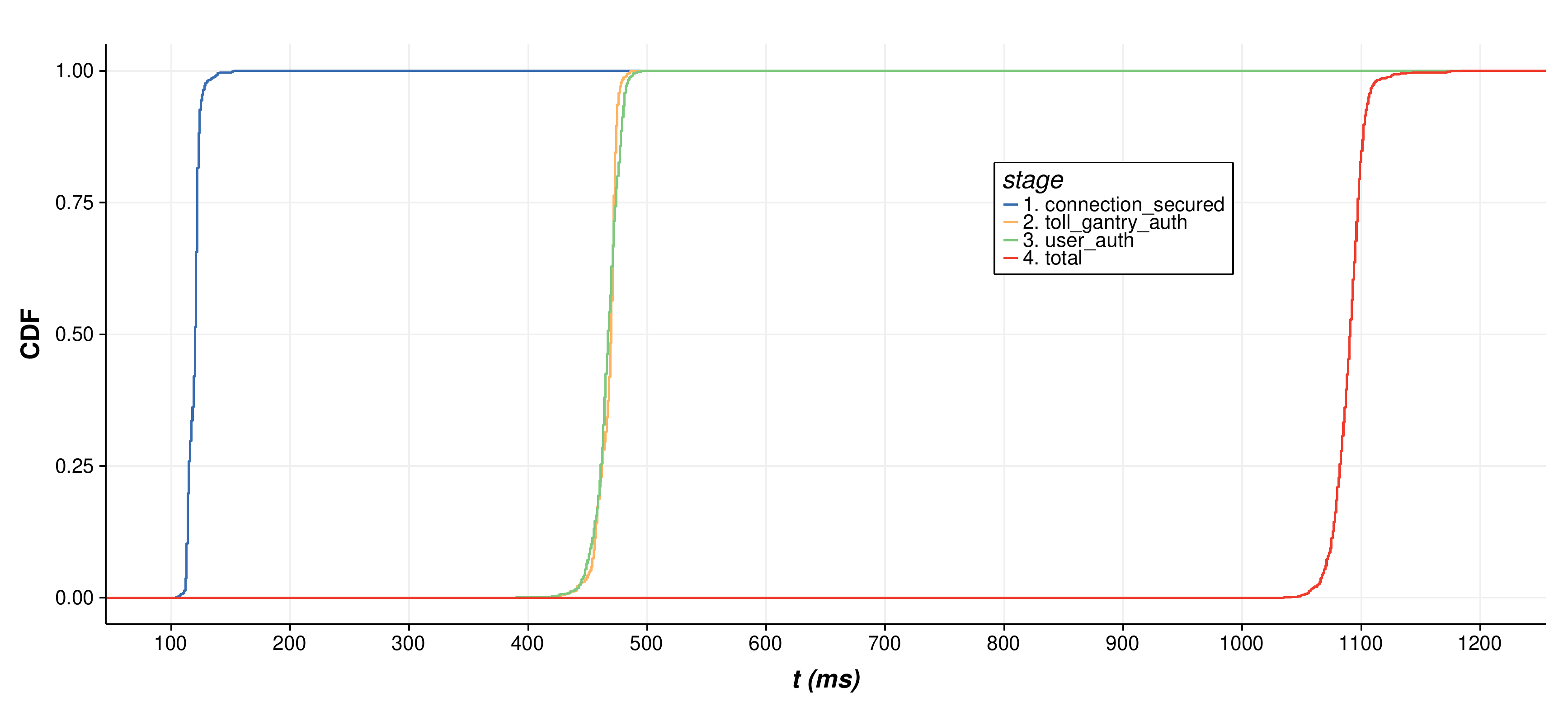}
        \caption{CDF of the latency experienced to establish a secure communication channel through the exchange of the on-spot created pairwise DIDs and both-ways authentication achieved through the exchange of on-spot created and validated proofs. }
        \label{fig:auth-ecdf}
\end{figure*}

The results of this assessment are documented in the Cumulative Distribution Function (CDF) illustrated in Fig. \ref{fig:auth-ecdf}. The average elapsed time for establishing a secure channel and validate the credentials of both parties (\textit{User} and \textit{Toll Gantry}) is 1090.3 ms. In average, establishing a secure pairwise DID connection took 119.4 ms, while the credential validation of the \textit{Toll Gantry} and \textit{User} lasted 467.0 ms and 465.9 ms, respectively.

An aspect that is worth mentioning is the latency determinism in all stages of the credential validation. The steep CDF latency curves indicate that the delays are concentrated around average values with little dispersion around them.

\subsection{IOTA Payment}

In order to evaluate the \emph{Tangle} transaction attachment latency, which represents the amount of time taken to conduct a toll payment, a test setup employing a personal computer, a IOTA node and a ``Proof-of-Work'' (PoW) accelerator encompassing an FPGA where used. The test setup is illustrated in Fig. \ref{fig:sub:iota}. The personal computer employs a i7-3630QM CPU, operating at 3.40 GHz with 8GB RAM memory.
The IOTA node is implemented on a four-core virtual machine running on a Xeon X5675 processor operating at 3.46 GHz with 12GB RAM memory. This node is connected to the IOTA network and runs IRI version 1.5.5 (IOTA's node software) while exposing its API through HTTP.
The PoW accelerator is a Cyclone 10 LP FPGA, named PiDiver \cite{pototschnig2018}, connected through its GPIO pins to a Raspberry Pi 3B running a HTTP server. In Fig. \ref{fig:sub:iota} the PiDiver plus Raspberry Pi is represented as a single entity named PiDiver.



Using IOTA's JavaScript library, \textit{iota.js}, value transactions (bundle of three transactions) were added to the \textit{Tangle} in order to simulate effective toll payments. This bundle includes in one of the transactions' message field (\textit{signatureMessageFragment}) the nonce related to the toll payment request issued by the \textit{Toll Gantry}. The bundle also holds information about the sender and receiver addresses as well as the value transferred between the two.

The process of attaching a value transaction to the \textit{Tangle} can be divided in three stages: selection of two transactions to be approved (tip selection), execution of the PoW (attach to Tangle), and network broadcast of the transaction (broadcast). The processing work related to these three stages were outsourced through HTTP requests by the personal computer to the other local network elements, the IOTA node and PiDiver.

\begin{figure*}[t]
	\centering
	\includegraphics[width=0.70\textwidth]{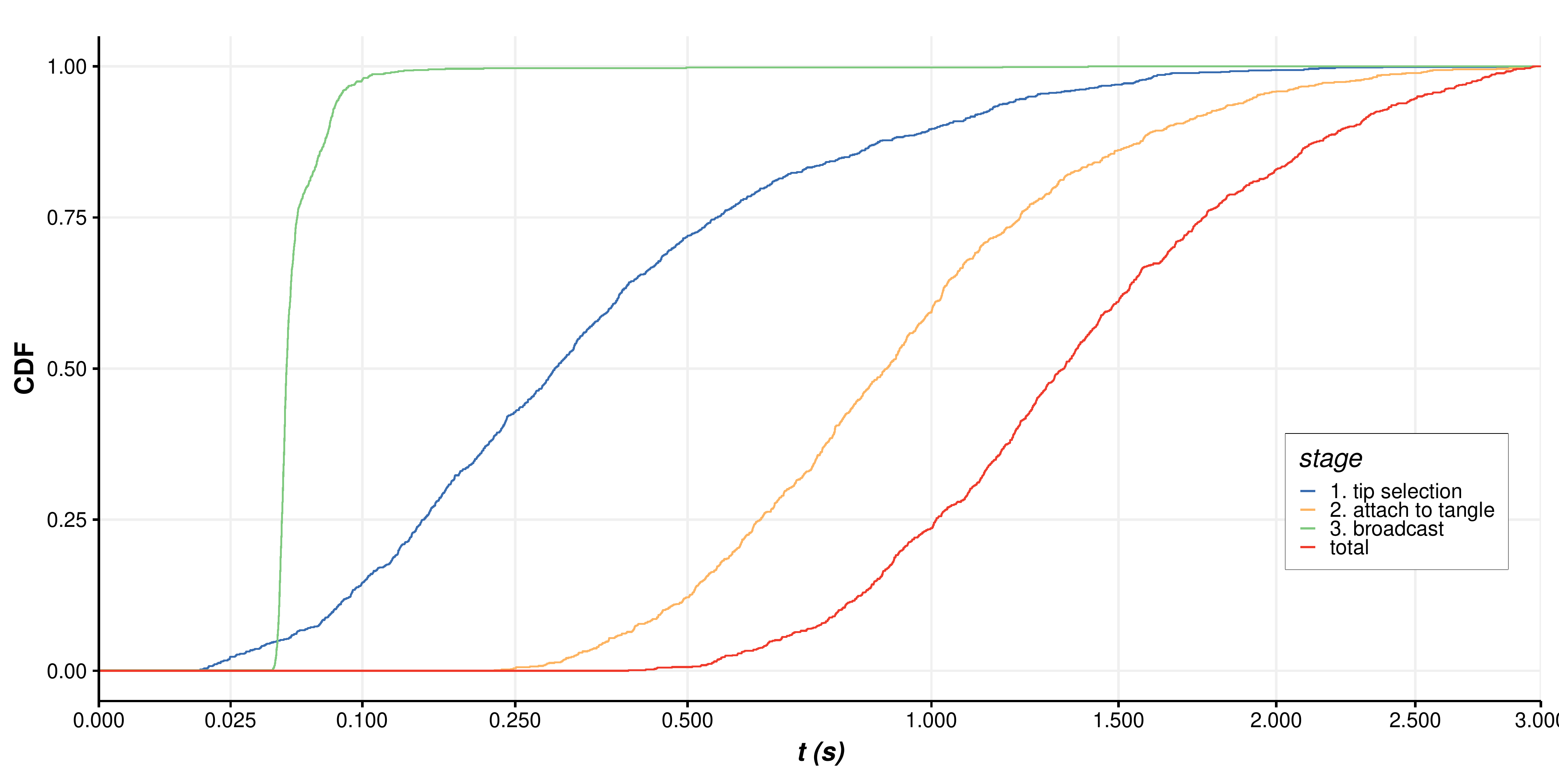}
        \caption{CDF of the latency experienced during the multiple phases of issuing a value transfer to the \textit{Tangle} (three signed transactions).}
        \label{fig:value-pidiver}
\end{figure*}

In the first stage, the tip selection algorithm is run two times, a process which was outsourced to the IOTA node as it requires an up-to-date ledger. 
In the second stage the proof of work is computed, a process which was outsourced to the PiDiver in order to shorten the time required to find a nonce. 
The third stage corresponds to the diffusion of the transaction in the \textit{Tangle}, a process which was outsourced to the IOTA node as it is connected to the IOTA network. 
The Cumulative Distribution Functions (CDF) of a 1000 iterations of the latency experienced in these three phases is plotted in Fig. \ref{fig:value-pidiver}.


As shown, the duration of the PoW has a high variance due to the randomness of finding a suitable nonce. The step with the higher impact in latency is the ``attach to tangle" component corresponding to the ``proof-of-work" that needs to be carried out for each transaction. The average latency is 1.02 s in this case. The second most significant contribution is the ``tip selection", averaging at 0.43 s. The message broadcast averages around 0.06 s. Overall, the global latency of adding a value transaction to the \textit{Tangle} using our setup is 1.51 seconds in average.


After a transaction is added to the \textit{Tangle}, and before it can be considered complete, it must be \textit{confirmed} in the \textit{Tangle}. Although this process is carried without the involvement of the  \textit{User} smartphone, it does implicate that the transaction cannot be deemed concluded until it is \textit{confirmed}. Indeed, the longer it takes for the transaction to be \textit{confirmed} the less probable it is to be. This is where \textit{reattach} and \textit{promote} techniques can be employed to foster its confirmation. This process is beyond the scope of this paper as it can be later executed by the \textit{Road Operator} or by the \textit{Toll Gantry}.

\subsection{Discussion}

Results for credential validation show that this process takes a short amount of time and it is highly deterministic when compared to attaching a \textit{Tangle} transaction. Although the amount of information used in the credential validation is adequate for the envisaged purpose, it may not fit other applications requiring more parameters in the credentials.

The results for the latency of a IOTA value transfer are, approximately, 40\% higher than those of the credential validation. To make that worse, a IOTA exchange among parties results not in one transaction added to the \emph{Tangle}, but in multiple transactions. These transactions are grouped into a so-called \textit{bundle} that represents an atomic transfer item in the \textit{Tangle}. The ``tip selection" and ``broadcast" are the same as for appending a single transaction. However, the PoW (``attach to tangle") must be carried out individually for each of the embedded transactions in the \textit{bundle}. At the recommended level of security for low-value transactions (level 2 - 256 bit signature) three transactions will be embedded in the bundle: debit, credit and debit signature. This number of transactions occurs when exchanging IOTAs only between two addresses/``accounts": one belonging to the \textit{User} and the the other to the \textit{Toll Gantry}. 

The overall latency of conducting the credential validation and performing the IOTA value exchange between \textit{User} and \textit{Toll Gantry} is demonstrated to last 2.6 seconds in average for the realized test setup. This result shows that an open road toll collection system solely based on the proposed architecture and operation can perform all steps within a bounded time window of 33.2 s, thus making the solution feasible and providing enough slack to handle higher vehicle speeds and support reduced communication ranges, for example.

As the trials were conducted in a very controlled communication environment, it is expected that common wireless phenomenons such as multi-path fading and interference will degrade the performance of the electronic tolling system, but without compromising its operation.

\section{Related work}
\label{rel-work}
Vehicle-to-everything (V2X) services will benefit from the enhanced performance of 5G systems, such as ultra-low latency, higher data rates, reliability and other features such as network slicing \cite{seaica}. Moreover, high speed data rate and low latency timings are expected to be accompanied with efficient authentication procedures. 

Legacy cellular networks provide a high level of security to its users. Traffic is encrypted and the user equipment (UE) and base station (BS) are mutually authenticated. However the new networking paradigms created by 5G advanced features require new security mechanisms.
In 5G, not only the UE and BS are authenticated mutually but the authenticity of  third party services needs to be verified \cite{s5gmwn}.

Traffic is expected to be largely offloaded to D2D connections between user equipment in order to decrease the computational burden on base stations. This allows a more efficient spectrum use but also gives rise to certain problems such as increased interference.

New attack vectors are expected to arise due to the more advanced services that 5G is capable of supporting. The multitude of services supported by 5G makes it very difficult to create a one-fits-all security architecture. Such architecture should be capable of sustaining DDoS and man-in-the-middle attacks, as well as jamming due to device interference and third party eavesdropping.

Vuk Marojevic \cite{vuk18}, discusses possible threat scenarios for C-V2X and concludes that, despite the safety-critical nature of C-V2X, only few mechanisms and procedures have be specified to secure it. In fact, the 3GPP R14 specifications for C-V2X explicitly state that no security is applied for the C-V2X broadcast type communication, i.e., messages exchanged among UEs have no standard security mechanisms in place. 

In \cite{lu20}, Lu \textit{et al}, present a comprehensive survey on the security of 5G V2X services together with an in-depth analysis of the state-of-the-art strategies for securing 5G V2X services and discussing how to achieve trust, security, or privacy protection in each strategy. 

According to Qualcomm \cite{qualcomm19}, as security over PC5 mode 4 relies on application layer security, there will be no difference in the privacy and security for the C-V2X when compared to IEEE 802.11p based solutions. This means that application-level security defined in IEEE WAVE (1609.2) and ETSI ITS G5 can be transparently used in C-V2X. Despite this possibility, for the use case consider in this paper, that would imply using a centralized public key infrastructure  \cite{ferfer} and a significant bandwidth penalty on the wireless medium \cite{ruf19}. 

Yang \textit{et al}, \cite{safeguarding}, discuss the process of securing 5G wireless communications using physical layer.
Physical layers security techniques use the characteristics of the wireless channel, multiple antennas, modulation and coding in order to avoid eavesdropping. These techniques have two major advantages: they don't require any considerable computational complexity, meaning that powerful unauthorized devices can't disturb the network security, and, they have a high scalability, avoiding the need of a sometimes difficult to implement cryptographic key distribution system where devices constantly join and leave network cells.

Key management in small network cell can be difficult where users constantly join and leave access points. The increased latency caused by the frequent handovers between cells can also impact the efficiency of the authentication processes. In order to leverage the benefits of SDN, a fast authentication method is presented in \cite{fa5ghn}. The proposed method is considered to be hard to be compromised due to the use of several physical layer attributes used as user fingerprints in the authentication process. User location is also predicted in order to prepare the next relevant cells. 

In \cite{spacavrs}, 5G is a potential candidate in vehicular networks (VANETs). A system architecture is proposed featuring real-time video services with emphasis on reliable, secure and privacy-aware communications. The architecture is composed by a mobile core network (MCN), a trusted authority (TA), a department of motor vehicles (DMV), and
a law enforcement agency (LEA). D2D and mmWave procedures are used in the communications.  The  proposed cryptographic mechanisms include a pseudonymous authentication scheme, a public key encryption with keyword search, a ciphertext-policy attribute-based encryption, and threshold schemes based on secret sharing.

In device-to-device (D2D) communications, devices can communicate with each other without going through base stations. A secure data sharing protocol (SeDS) for D2D communication in 4G LTE-Advanced is proposed in \cite{seds}. The proposed method achieves various security requirements through the use of digital signatures and symmetric encryption without adding extra load to the network. The proposed method can detect free-riding attacks by keeping records of the current user equipment in the network.

In \cite{secure} the authors propose a secret key sharing method between two devices without the need of prior knowledge between both of them in order to achieve secure D2D communications. This is achievable through a low computational cost and small mutual authentication overhead. The proposed method was implemented using two smartphones and the Wi-Fi Direct protocol, demonstrating the efficiency and usability of the proposal.

The physical layer can also be used to secure D2D communications. In \cite{physical}, the authors propose the use of a beam forming technique that uses physical layer coding to secure information as it passes through a relay. Results show that the algorithm converges fast and D2D performance is degraded as the number of eavesdroppers rise.

D2D nodes in proximity can share sensitive information relative to their users identity and personal details \cite{5gd2dn}. A eavesdropper could exploit D2D communication weaknesses and as therefore use the stolen information for illegal purposes. Contributing to D2D security, in \cite{sd2dwl}, the concept of continuous authenticity and a security scoring system (SeS) for measuring security is proposed. Legitimacy patterns, which are sequence of bits inserted in the transmitted packets continuously are introduced to give an advantage over the attackers. In order to measure the security, the security scoring system is based on the violation of the legitimacy patterns. Simulation results show an efficient detection of attacks without requiring intensive computations at higher layers of the software stack.

\section{Conclusions}
\label{conclusions}

The deployment of 5G C-V2X technology will foster a richer service ecosystem for vehicular applications. The emergence of new cryptography based technologies for digital identity and currency will stem new solutions that will meet existing and future challenges. This paper proposes a crypto tolling architecture that harnesses 5G C-V2X connectivity between \textit{User} and \textit{Toll Gantry} devices in order to establish a secure, private and convenient tolling system. Besides describing the architecture and operation of the envisaged solution, the paper also documents a feasibility analysis conducted using an experimental setup. Results show that an open road tolling system can be realized using the IOTA framework for payments and the Hyperledger Indy for self-sovereign identity validation. Results also show that the use of dedicated GPU hardware can  ensure the execution of IOTA's PoW in a timely fashion.




%

\begin{IEEEbiography}[{\includegraphics[width=\textwidth]{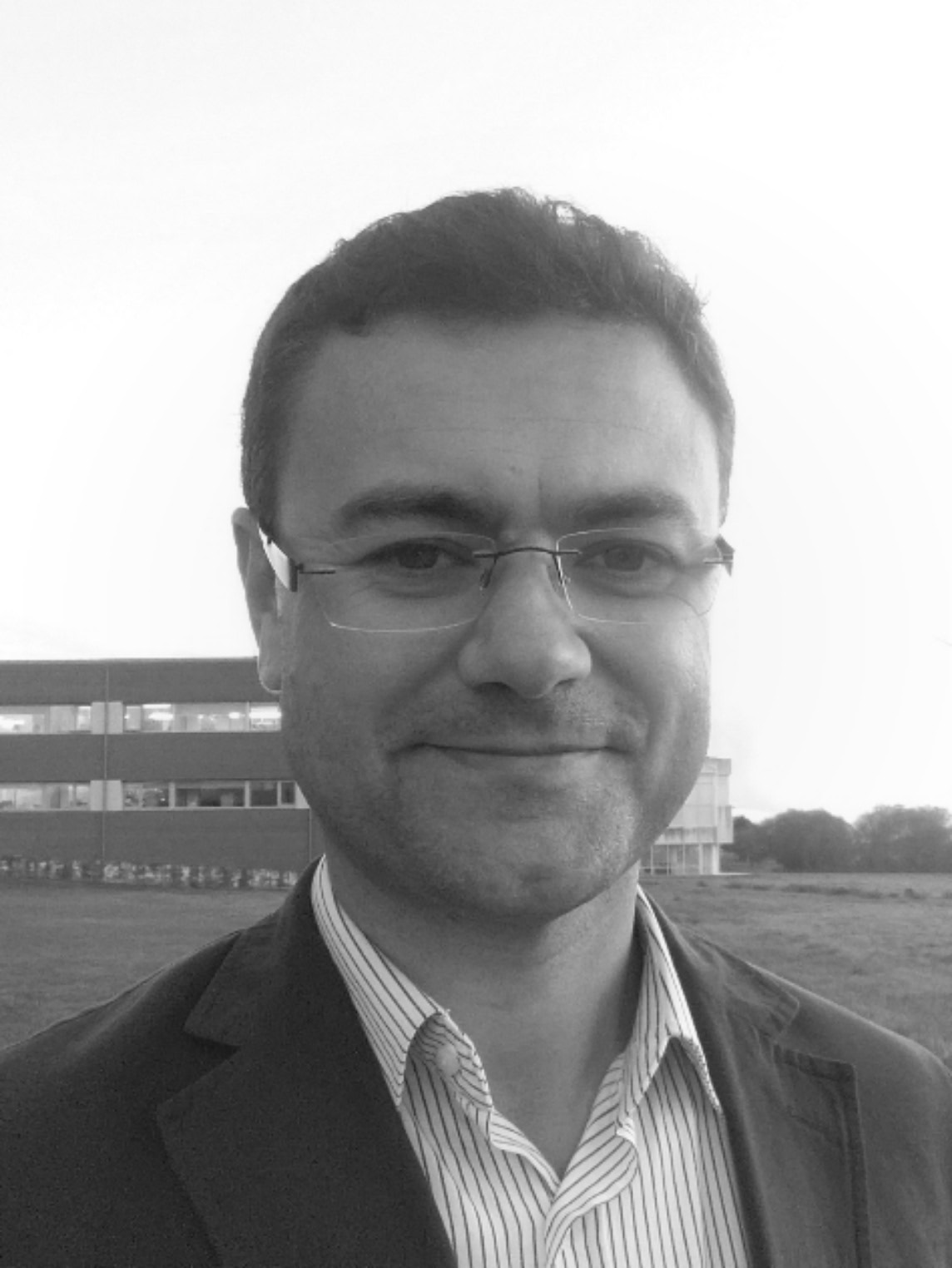}}]{Paulo C. Bartolomeu}
received his Ph.D. in Informatics Engineering from the University of Aveiro, Portugal, in 2014. He has participated in several R\&D projects both at the academia (ARMONIO, CAMBADA) and in the industry (CIRaF, DHT-Mesh, BikeEmotion, LUL, SheepIT). He is the author of two patents and more than 40 scientific publications including papers in conferences, journals and book chapters. His research interests include real-time communications, information centric networks, blockchain and IoT.
\end{IEEEbiography}

\begin{IEEEbiography}[{\includegraphics[width=\textwidth]{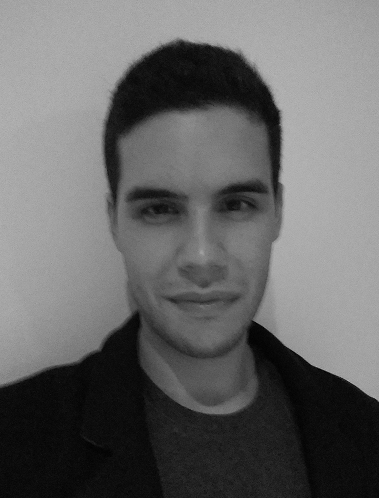}}]{Emanuel Vieira}
received his Master's degree in Engineering Physics from the University of Aveiro in Portugal, in 2017. His research interests include machine learning and blockchain technologies.
\end{IEEEbiography}


\begin{IEEEbiography}[{\includegraphics[width=\textwidth]{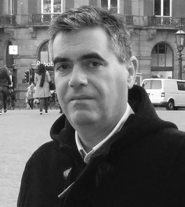}}]{Joaquim Ferreira}
received Ph.D. degree in Informatics Engineering from University of Aveiro, Portugal in 2005. Currently, he is an adjunct professor at School of Technology and Management, University of Aveiro and researcher at Telecommunications Institute.  
He has been involved in several international and national research projects. His research interests include: dependable distributed systems, fault-tolerant real-time communications, wireless vehicular communications, cooperative ITS systems and medium access control protocols. He is the author of several scientific papers and book chapters in his areas of expertise. He is senior member of IEEE and served in several conference scientific committees.
\end{IEEEbiography}





\begin{thebibliography}{99}

\bibitem{Persad07}
Khali Persad, C. Michael Walton, Shahriyar Hussain, Toll Collection Technology and Best Practices, Techichal Report 0-5217-P1, Center for Transportation Research, The University of Texas at Austin, 2007.  

\bibitem{vlachos2017} C. Vlachos and V. Friderikos, "MOCA: Multi objective Cell Association for Device-to-Device Communications," in IEEE Transactions on Vehicular Technology, vol. 66, no. 10, pp. 9313-9327, Oct. 2017. 

\bibitem{cv2x}
3GPP, “TS 36.300 E-UTRA and E-UTRAN; Overall description; Stage 2 (v14.8.0, Release 14),” 3GPP, Tech. Rep., Oct. 2018. 

\bibitem{e1} 
S. Popov,  "The \emph{Tangle}", IOTA Whitepaper version 1.3, October 2017, http://iotatoken.com/IOTA\_Whitepaper.pdf [Online].  Accessed in 11/2/2020.


\bibitem{sovrinwhitepaper} Sovrin Foundation,  "Sovrin™: A Protocol
and Token for Self-Sovereign Identity and Decentralized Trust", Sovrin Whitepaper version 1.0, January 2018, https://sovrin.org/library/sovrin-protocol-and-token-white-paper/ [Online].  Accessed in 26/01/2020.

\bibitem{5gaa}
5GAA Automotive Association, "Addendum to V2X Functional and Performance Test Report; Test Procedures and Results for 20-MHz Deployment in CH183", Tech. Rep., May 2019.

\bibitem{pototschnig2018}  Thomas Pototschnig,  ``IOTA PoW Hardware Accelerator FPGA for Raspberry Pi (und USB)", April 2018, https://microengineer.eu/2018/04/25/iota-pearl-diver-fpga/ [Online]. Accessed in 26/01/2020.

\bibitem{seaica}
Muhammad, M., and Safdar, G. A. (2018). Survey on existing authentication issues for cellular-assisted V2X communication. \textit{Vehicular Communications, 12}, 50-65. doi:10.1016/j.vehcom.2018.01.008

\bibitem{s5gmwn}
Fang, D., Qian, Y., and Hu, R. Q. (2018). Security for 5G Mobile Wireless Networks. \textit{IEEE Access, 6}, 4850-4874. doi:10.1109/access.2017.2779146

\bibitem{vuk18}
Vuk Marojevic,"C-V2X Security Requirements and Procedures: Survey and Research Directions", arXiv:1807.09338, 2018.

\bibitem{lu20}
R. Lu, L. Zhang, J. Ni and Y. Fang, "5G Vehicle-to-Everything Services: Gearing Up for Security and Privacy," in Proceedings of the IEEE, vol. 108, no. 2, pp. 373-389, Feb. 2020.

\bibitem{qualcomm19}
Qualcomm, "C-V2X Technical Performance Frequently Asked Questions", 80-PE732-67 Rev. A, Tech. Rep., Oct., 2019.

\bibitem{ferfer} 
B.  Fernandes,  J.  Rufino,  M.  Alam,  and  J.  Ferreira,  “Implementation and  Analysis  of  IEEE  and  ETSI  Security  Standards  for  Vehicular Communications,” Mobile   Networks   and   Applications,   Feb   2018. [Online]. Available: https://doi.org/10.1007/s11036-018-1019-x

\bibitem{ruf19}
J. Rufino, L. Silva, B. Fernandes, J. Almeida and J. Ferreira, "Overhead of V2X Secured Messages: An Analysis," 2019 IEEE 89th Vehicular Technology Conference (VTC2019-Spring), Kuala Lumpur, Malaysia, 2019, pp. 1-5. doi: 10.1109/VTCSpring.2019.8746479


\bibitem{safeguarding}
Yang, N., Wang, L., Geraci, G., Elkashlan, M., Yuan, J., and Renzo, M. D. (2015). Safeguarding 5G wireless communication networks using physical layer security. \textit{IEEE Communications Magazine}, 53(4), 20-27. doi:10.1109/mcom.2015.7081071

\bibitem{fa5ghn}
Duan, X., and Wang, X. (2016). Fast authentication in 5G HetNet through SDN enabled weighted secure-context-information transfer. \textit{2016 IEEE International Conference on Communications (ICC)}. doi:10.1109/icc.2016.7510994

\bibitem{spacavrs}
Eiza, M. H., Ni, Q., and Shi, Q. (2016). Secure and Privacy-Aware Cloud-Assisted Video Reporting Service in 5G-Enabled Vehicular Networks. \textit{IEEE Transactions on Vehicular Technology, 65}(10), 7868-7881. doi:10.1109/tvt.2016.2541862

\bibitem{seds}
A. Zhang, J. Chen, R. Q. Hu and Y. Qian, SeDS: Secure Data Sharing Strategy for D2D Communication in LTE-Advanced Networks, \textit{IEEE Transactions on Vehicular Technology}, vol. 65, no. 4, pp. 2659-2672, April 2016. doi: 10.1109/TVT.2015.2416002

\bibitem{secure}
Shen, W., Hong, W., Cao, X., Yin, B., Shila, D. M., and Cheng, Y. (2014). Secure key establishment for Device-to-Device communications. \textit{2014 IEEE Global Communications Conference}. doi:10.1109/glocom.2014.7036830

\bibitem{physical}
Jayasinghe, K., Jayasinghe, P., Rajatheva, N., and Latva-Aho, M. (2015). Physical layer security for relay assisted MIMO D2D communication. \textit{2015 IEEE International Conference on Communication Workshop (ICCW)}. doi:10.1109/iccw.2015.7247255

\bibitem{5gd2dn}
Ansari, R. I., Chrysostomou, C., Hassan, S. A., Guizani, M., Mumtaz, S., Rodriguez, J., and Rodrigues, J. J. (2018). 5G D2D Networks: Techniques, Challenges, and Future Prospects. \textit{IEEE Systems Journal, 12}(4), 3970-3984. doi:10.1109/jsyst.2017.2773633

\bibitem{sd2dwl}
Abualhaol, I., and Muegge, S. (2016). Securing D2D Wireless Links by Continuous Authenticity with Legitimacy Patterns. \textit{2016 49th Hawaii International Conference on System Sciences (HICSS)}. doi:10.1109/hicss.2016.713





















































\end{thebibliography}
\end{document}